\documentclass[12pt]{article}

\usepackage{epsf,cite,a4wide,latexsym}
\newcommand{\beq}{\begin{equation}}
\newcommand{\eeq}{\end{equation}}
\newcommand{\bea}{\begin{eqnarray}}
\newcommand{\eea}{\end{eqnarray}}
\newcommand{\bda}{\begin{eqnarray*}}
\newcommand{\eda}{\end{eqnarray*}}

\newcommand{\N} {{\rm N}}

\begin{document}

\vskip -4cm

\begin{flushright}
MIT-CTP-2967
\end{flushright}

\vskip 0.2cm

{\Large
\centerline{{\bf  Vortex configurations in the large N limit.}}
\vskip 0.3cm

\centerline{ \'Alvaro Montero }}
\vskip 0.3cm

\centerline{Center for Theoretical Physics}
\vskip -0.15cm
\centerline{Laboratory for Nuclear Science and Department of Physics}
\vskip -0.15cm
\centerline{Massachusetts Institute of Technology}
\vskip -0.15cm
\centerline{Cambridge, Massachusetts 02139}
\vskip -0.15cm
\centerline{USA}
\vskip 0.1cm
\centerline{e-mail: montero@mitlns.mit.edu}
\vskip 0.8cm

\begin{center}
{\bf ABSTRACT}
\end{center}
We study the properties of vortex-like configurations which are solutions 
of the SU(N) Yang-Mills classical equations of motion. We show that
these solutions are concentrated along a two-dimensional wall with size
growing with the number of colors.  

\vskip 1.5 cm
\begin{flushleft}

\end{flushleft}

\newpage

\section{Introduction.}

The vortex condensation theory is one of the most promising ideas to explain the
property of confinement. This idea was proposed by many authors at the end of 70's
\cite{thooft} and has received a renewed interest in a set of recent works 
\cite{green1,green2,green3,green4,green5,green6,lang,lang2,lang3,cher,forcrand,forcrand2,kovacs}.
A gauge dependent approach is used in 
\cite{green1,green2,green3,green4,green5,green6,lang,lang2,lang3,cher,forcrand,forcrand2} 
to study the relation between the confinement property and the vortex 
contain of the vacuum.  By first going to the maximal center gauge and then center projecting, 
it is found evidence of a center vortex origin 
for the confinement property. In references \cite{kovacs} is presented another approach
to the same problem, in this case relating the vortex contain of the vacuum and
confinement in a gauge invariant way.

One of the problems of this idea is that it can not explain the behaviour 
of the string tension in the adjoint representation.  
This quantity shows casimir scaling for small distances 
and vanishes for large distances. Casimir scaling is a property
of the ratio between the string tension in the adjoint 
representation, $\sigma_{A}$, and the string tension in the fundamental
one, $\sigma_{F}$. This quantity $\sigma_{A}/\sigma_{F}$ is
equal to the ratio between the casimirs of the two representations in the
small distance region \cite{kov}. In the large distance region the adjoint
string breaks because quarks in the adjoint representation and gluons can
mix to built a bound state. Nevertheless, in the limit of large number
of colors it is expected that the property of confinement also holds. 
This fact is expected because of the factorization property for gauge
invariant operators, which holds in the large N limit. 
The expected value of a Wilson loop along a path $C$ calculated in the adjoint representation 
$ \langle W_{A}(C) \rangle$ can be written as $\langle W_{A}(C) \rangle = 
\langle W_{F}(C) W_{F}(C)^{\dag} \rangle$, being $W_{F}(C)$ the loop in 
the fundamental representation, and, using the factorization property, 
we obtain that $ \langle W_{A}(C) \rangle = \langle W_{F}(C)\rangle^2$. 
Then, in the large N limit it is expected that the adjoint string tension 
is twice the fundamental one.
These two things, casimir scaling of the string tension and confinement of
adjoint quarks in the large N limit, could be in contradiction with the 
model of confinement based in physical objects carrying 
flux quantized in elements of the center of the group, because the adjoint 
representation is blind to these elements. 
The proposal presented in reference \cite{green2} to solve this
problem is that the approximate casimir scaling is an effect due to the 
finite thickness of the vortex configuration, and also that the non vanishing
adjoint string tension in the large N limit can be explained if this 
thickness grows with the number of colors. Following this idea, we study the
behaviour with the number of colors of the vortex solution presented in
\cite{R2xT2,CGF} for the SU(2) and SU(3) groups, and we found that for 
$N>3$ the solution has the same properties and his size grows with the number 
of colors, N.

The paper is structured as follows. In section 2 we show the properties 
of the solution we have found and in section 3 we present our conclusions.

\section{The solutions.}
Following the same methods used in \cite{cool1,R2xT2,CGF} we isolate solutions
of the SU(N) Yang-Mills classical equations of motion on a four torus with sizes  
${l_L}^2 \times {l_S}^2$, being $l_L \gg l_S$ (two large, $t$ and $x$, 
and two small, $y$ and $z$, directions).
These solutions satisfy twisted boundary conditions given by the twist vectors
$\vec k = \vec m = (1,0,0) $, have topological charge $|Q|=1/N$ and action 
$S=8\pi^2/N=8\pi^2|Q|$ (a review of the properties of Yang-Mills fields on
the torus can be found in \cite{tonyrev}). The choice of the twist vectors is based in the results
presented in reference \cite{R2xT2}. In this article it is shown that the necessary
condition to obtain a SU(2) solution which is localized in the two large directions,
flat in the other two and having the properties of a vortex is that the twist in the
small plane must be non trivial. The same property also holds for SU(3) \cite{CGF} 
and we expect the same result for SU(N) groups, with $N > 3$. For this reason we
choose $m_1 = 1$, because this is the component of the twist linked with the small 
plane, the $yz$ plane.  

We have isolated lattice configurations for values of the number of colors $N=4,5,6,7,$
$8,9,10$. As in \cite{cool1,R2xT2,CGF} these configurations are obtained using standard 
cooling algorithms (our procedure is a local minimization of the Wilson action
using the Cabbibo-Marinari-Okawa update \cite{Cabbibo}). 
Our criteria to stop the cooling procedure is that the Wilson action 
is stable up to the eight significant digit and close to
the value $S=8\pi^2/N$. In table 1 we give the set of studied solutions, specifying
the number of colors, lattice size $(N_L)^2 \times (N_S)^2$ 
and the values of the action S (in $8\pi^2/N$ units), 
topological charge Q multiplied by N and the electric and magnetic parts of the action, 
$S_e$ and $S_b$ respectively (also in $8\pi^2/N$ units). These quantities are
calculated from the field strength ${\bf F}_{\mu\nu}$ obtained from the clover average
of plaquettes $1 \times 1$ and $2 \times 2$, combined in such a way that the 
discretization errors are $O(a^4)$. The continuum expected values for these quantities
are: $SN/8\pi^2=1$, $QN=\pm1$, $S_e/8\pi^2=0.5$ and $S_b/8\pi^2=0.5$. We observe
that our lattice results are very near to the continuum values and also that the lattice
corrections are smaller for larger values of N.

The choice of the sizes of the lattice is based in references \cite{R2xT2,CGF}. 
From the work \cite{R2xT2} we learn that for the SU(2) case the obtained
results show a very nice scaling towards continuum functions for lattice sizes 
$(4N_S)^2\times(N_S)^2$ with $N_S = 4,5,6,7$. The SU(3) case is studied in \cite{CGF}
and it is shown in this reference that for lattice sizes $(6N_S)^2\times(N_S)^2$ with
$N_S = 4,5,6$ it is also found a very nice scaling towards continuum functions. Following
these results we use lattices with sizes $(2NN_S)^2\times(N_S)^2$. 
For number of colors $N=4$ we isolate solutions with $N_S = 2,3,4$ and study 
their scaling properties. We observe that with only two points in the small directions
the obtained results show a very nice continuum behaviour (to set the scale 
we fix the length of the small direction, $l_S$, equal to $1$, 
being therefore the lattice spacing $a=1/N_S$). 
We illustrate this in figure 1 in which we show the energy profile $\epsilon(t)$, 
defined as,
\beq
   \epsilon(t) = \int S(t,x,y,z) dx dy dz  \label{eq:enprof}, 
\eeq
where $S(t,x,y,z)$ is the action density, for these three SU(4) configurations.
This is a property which also holds for bigger values of N and allows us to use
lattices with sizes $N_S=2$ and still obtain good continuum results. In fact,
we can use this lattice size because these solutions are almost flat in the
small directions and then very few points are needed to describe it.

Now, we describe the shape and the N behaviour of the action density and the 
integrated quantity defined before, the energy profile. We start with the N 
behaviour of the energy profile. The result we get is that this
function has a typical width which grows with the numbers of colors as the 
square root of this number. We illustrate this property in figure 2 in which 
we plot $\epsilon(t)$ multiplied by $N^{3/2}$
as a function of $\sqrt{N}$. From this figure we can clearly see that the size
of the solution grows with the square root of the number of colors, and also
the scaling of the energy profile with $N^{3/2}$. 

The properties of the action density  $S(t,x,y,z)$ for our solutions are the
following. The function $S$ is almost independent of the $y$ and $z$ coordinates,
and has a instanton-like dependence in the $t$ and $x$ coordinates. The function
in these coordinates, $t$ and $x$, has only one maximum, that we put at point 
$t=x=0$, and decreases with $|t|$ and $|x|$ up to the value $S=0$, showing an 
exponential behaviour at the tails. The size of the solution in these two
directions, $t$ and $x$, is approximately the square root of the number of 
colors (remember that the length scale is set by the size of the small 
direction $l_S=1$). 

The main result of this work is that these solutions have vortex properties for
all the values of N studied and that the size of the object grows with the number 
of colors. To illustrate these properties we calculate the Wilson loop $W_C(r)$ 
around this object, being the path C a $r \times r$ square loop in the x,t plane
centered at the maximum of the action density. We parametrize $W_C(r)$ by the 
functions $L(r)$ (its module) and $\phi (r)$ (its phase). Now we describe the
behaviour of these two functions with the size of the loop r.
The module $L$ takes the value $L=1$ at point $r=0$, decreases with $r$ up to reach
the minimum of the function,  and then increases up to the value $L \sim 1$ at a point $r$
between $\sqrt{N}$ and $N$.
The phase $\phi$ takes the value $\phi=2 \pi $ at $r=0$ and then decreases up to
the value $\phi=2 \pi (1-\frac{1}{N})$ at a point $r$ between $\sqrt{N}$ 
and $N$. Then, for a enough big 
value of $r$ the Wilson Loop $W_C(r)$ takes the value of an element of the center
of the group, $W_C \rightarrow exp(-i2\pi/N)$. We illustrate this property
in figures 3 and 4. In figure 3 we show the phase $\phi$ for the three SU(4) solutions 
as a function of the size of the loop r. First, we can see the nice scaling towards
a continuum function of the points coming from different lattice sizes, $N_s=2,3,4$.
And second, how the phase $\phi$ rotate from the value $\phi=2 \pi$ at point $r=0$
to the value $\phi= 3\pi/2$ at point $r \sim N=4$. Finally, we study the N 
behaviour of the phase $\phi$. In figure 4 we show the quantity
$\Phi=(2\pi-\phi)N$ as a function of $r/\sqrt{N}$ for the solutions with $N=4,7,10$. 
If the Wilson loop $W_C(r)$ takes the value $W_C \rightarrow exp(-i2\pi/N)$ for a enough
big value of $r$ then this phase $\Phi$ must take the value $\Phi=2\pi$ 
for the same value of $r$.
From this figure we can clearly see that this phase $\Phi$ reach the value $\Phi=2\pi$, and
that the value of $r$ in which this is happening, grows with the number of colors with a 
tendency a bit faster than the square root of the number of colors.

\section{Conclusions.}   

In this work we have shown that there are solutions of the SU(N) Yang-Mills equations of
motion having vortex properties and a size growing with the number of colors. These
solutions live on a four torus $T^4$ with sizes ${l_L}^2 \times {l_S}^2$, being 
$l_L \gg l_S$, and satisfy twisted boundary conditions. Looking at the action density
we have seen that are flat in the small directions and are localized in the 
two large directions, having a size which grows as the square root of
the number of colors N. We calculate a square Wilson loop in the plane formed by
the two large directions centered at the maximum of the solution, and observe
that for a enough big size of the loop the value obtained is an element of
the center of the group, $exp(-i2\pi/N)$. 

Note that, if we consider the limit $l_L \rightarrow \infty$ , we can say that
the solutions live on $R^2 \times T^2$. And further on, if we repeat the solutions
in the small directions , we can say that we have a solution of the SU(N) Yang-Mills 
equations of motion in $R^4$. In this case the solution will have infinite action
and infinite topological charge, will be flat in two of the coordinates
and localized in the other two, in a region of size approximately equal to the 
square root of the number of colors.

The relevant point of these solutions is that they are presenting the expected 
behaviour described in \cite{green2} for a confining configuration. They are bidimensional
objects carrying a flux quantized in elements of the center of the group, and
have a thickness which grows with the number of colors. Remember that, as we 
mention in the introduction, this finite thickness may be the explanation for
the casimir scaling of the string tension, and also, that the growing of this
thickness with the number of colors may take into account the confinement property
of quarks in the adjoint representation which holds in the limit of large number
of colors. 

Another interesting point is the one presented in reference \cite{CGF}.
After going to the maximal center gauge and center projecting the configuration,
the vortex solution for the SU(2) and SU(3) group appears as a projected vortex, 
the objects to which is attributed the confinement property. We expect that
the same property holds for group SU(N) with $N>3$, and then, these solutions
will be identified as projected vortices after going to maximal center gauge
and center projecting the configuration.

Finally, we want to say that our claim in this paper is not that our solution
is the basic ingredient to explain the confinement property. Nevertheless, we think
that this kind of configurations may play a role in the confinement mechanism, and,
in any model based in a bunch of classical solutions as, for example, the instanton 
liquid model, these solutions must be included. This kind of model has been presented
in \cite{TP} and some favourable results have been reported in \cite{TA1,TA2}. In this
model the object which was included is also a fractional charge solution, 
in this case having point-like properties.

\section*{Acknowledgements}
I acknowledge useful discussions with Margarita Garc\'{\i}a P\'erez, 
Antonio Gonz\'alez-Arroyo, John Negele, Carlos Pena and Manuel P\'erez-Victoria. 
This work has been supported by the spanish Ministerio de Educaci\'on y Cultura 
under a postdoctoral Fellowship.

\newpage

\begin{table}
\begin{center}
\vspace{-0.5 cm}
\caption{{\footnotesize Set of studied solutions. We specify the number of colors N,
the lattice size $(N_L)^2 \times (N_S)^2$ and the values of the action $S$ (in $8\pi^2/N$ units),
topological charge multiplied by N, electric part of the action $S_e$ and magnetic part of the
action $S_b$ (both in $8\pi^2/N$ units).  } }
\vspace{0.2 cm}
\begin{tabular}{||c|c||c|c|c|c||}
\hline
N  &  $(N_L)^2 \times (N_S)^2$        & $\frac{S \N}{8\pi^2}$ 
   &  Q \N                            & $ \frac{S_e \N}{8\pi^2}$  
   &  $ \frac{S_b \N}{8\pi^2}$  \\   
\hline \hline
4  &  $16^2 \times 2^2 $  & 0.96074  & -0.95659  &  0.49696  &   0.46378   \\ \hline
4  &  $24^2 \times 3^2 $  & 0.99643  & -0.99603  &  0.50412  &   0.49231   \\ \hline
4  &  $32^2 \times 4^2 $  & 0.99934  & -0.99921  &  0.50313  &   0.49621   \\ \hline
5  &  $20^2 \times 2^2 $  & 0.98237  & -0.98053  &  0.50211  &   0.48026   \\ \hline
5  &  $30^2 \times 3^2 $  & 0.99851  & -0.99826  &  0.50392  &   0.49459   \\ \hline
6  &  $24^2 \times 2^2 $  & 0.99061  & -0.98957  &  0.50380  &   0.48680   \\ \hline
6  &  $36^2 \times 3^2 $  & 0.99930  & -0.99912  &  0.50351  &   0.49579   \\ \hline
7  &  $28^2 \times 2^2 $  & 0.99450  & -0.99382  &  0.50439  &   0.49011   \\ \hline
8  &  $32^2 \times 2^2 $  & 0.99653  & -0.99605  &  0.50445  &   0.49209   \\ \hline
9  &  $36^2 \times 2^2 $  & 0.99770  & -0.99733  &  0.50439  &   0.49331   \\ \hline
10 &  $40^2 \times 2^2 $  & 0.99843  & -0.99813  &  0.50420  &   0.49422   \\ \hline
\end{tabular}
\end{center}
\label{tb:ressol}
\end{table}

\newpage

\begin{figure}
 \caption{The energy profiles $\epsilon(t)$ for the three SU(4) solutions, each one obtained
           on a lattice of size $ (8 N_s)^2 \times (N_s)^2$, are shown as a function of t.  }
 \vbox{ \vskip -1.5 cm \hskip -0.3 cm \hbox{  \epsfxsize=450pt \hbox{\epsffile{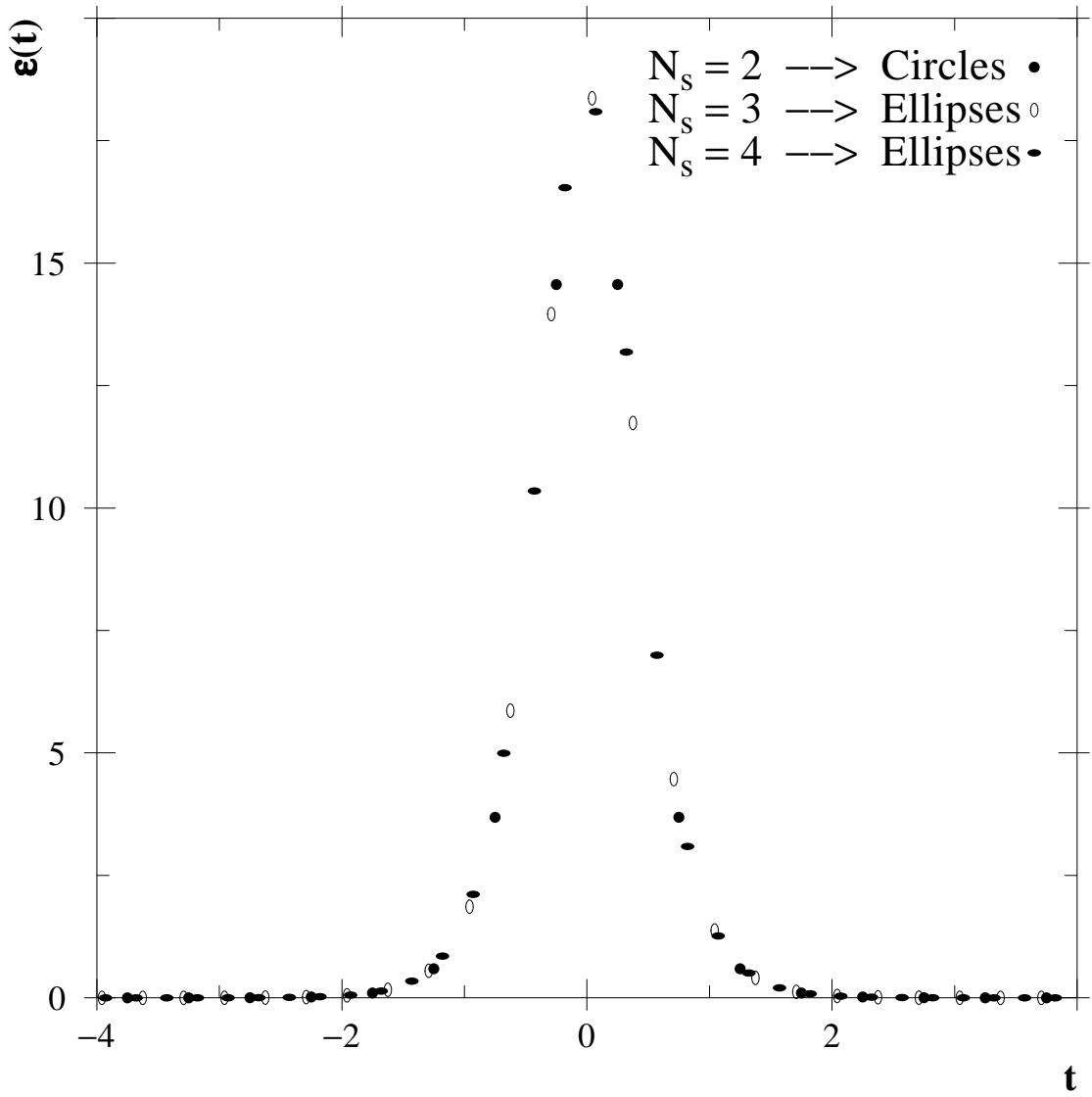} } } } 
\end{figure}

\newpage

\begin{figure}
 \caption{ The energy profiles $\epsilon(t)$ multiplied by $N^{3/2}$ for the solutions with
           N=4,7,10 are shown as a function of $t/N^{1/2}$. }
 \vbox{ \vskip -1.5 cm \hskip -0.3 cm \hbox{  \epsfxsize=450pt \hbox{\epsffile{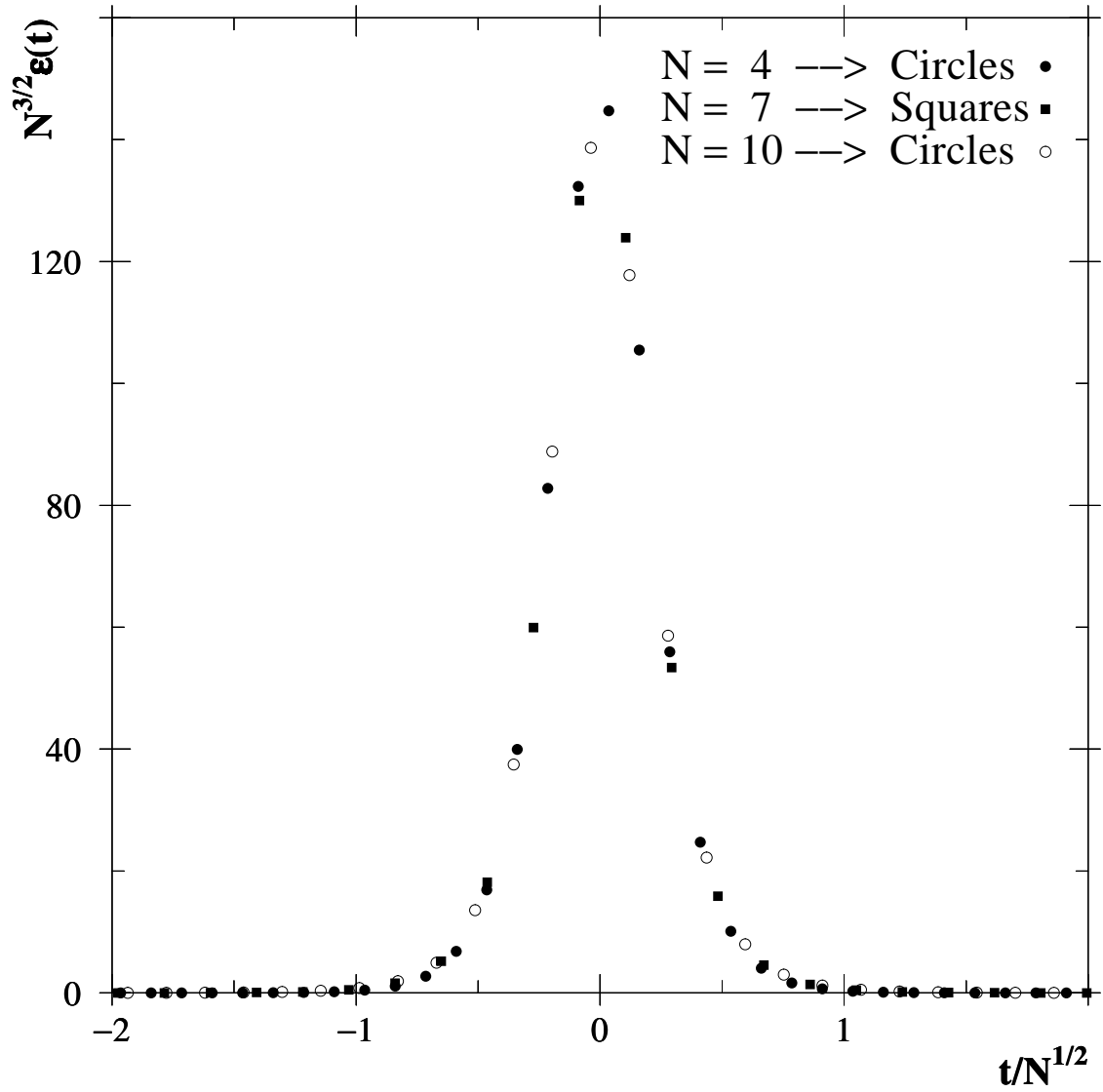} } } } 
\end{figure}

\newpage

\begin{figure}
 \caption{ The phase $\phi(r)$ of the Wilson loop $W_C(r)$ is shown as a function of 
           $r$ for the three SU(4) solutions, each one obtained on a lattice with size
           $(8 N_s)^2 \times (N_s)^2$.}
 \vbox{ \vskip -1.5 cm \hskip -0.3 cm \hbox{  \epsfxsize=450pt \hbox{\epsffile{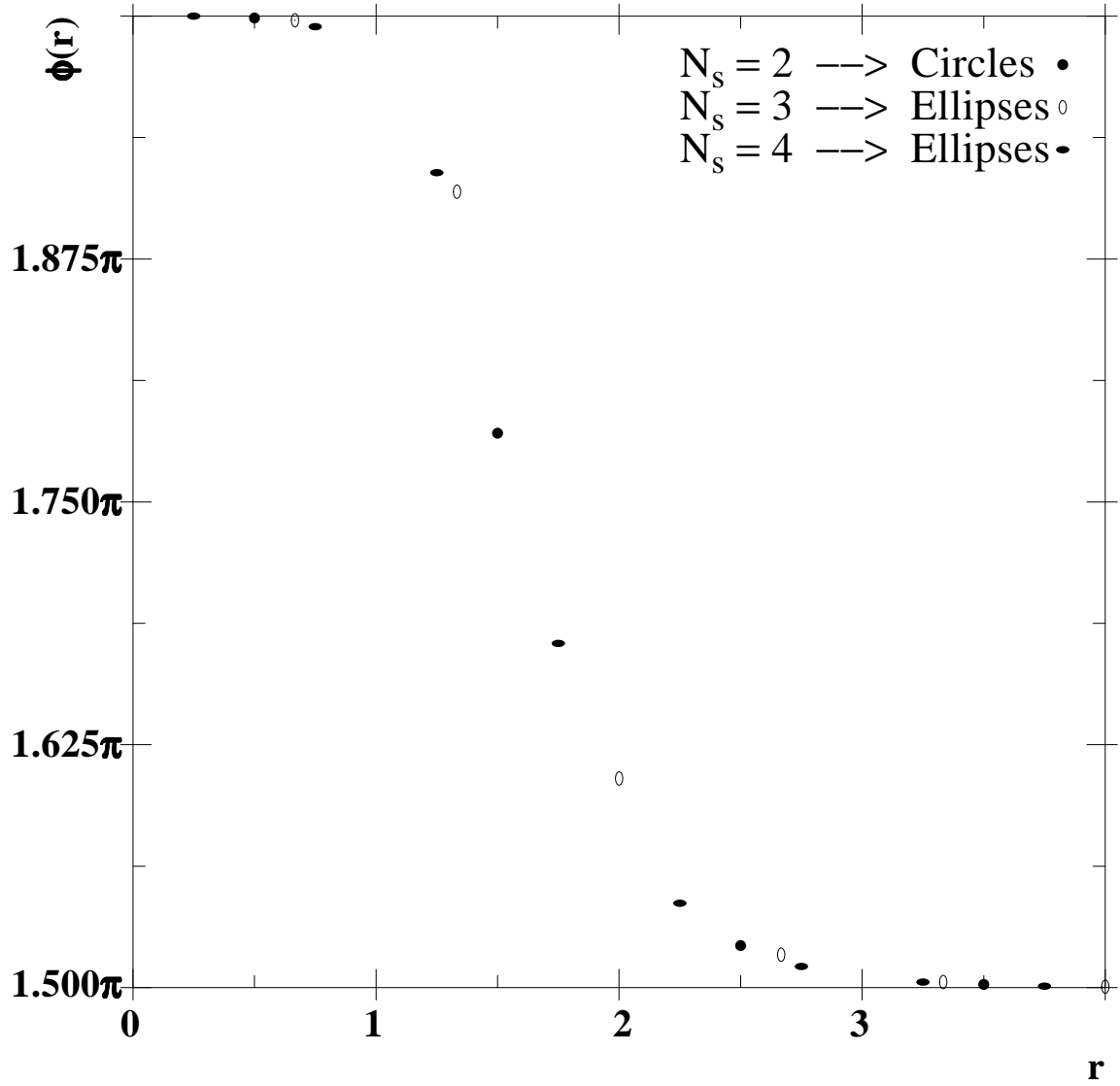} } } } 
\end{figure}

\newpage

\begin{figure}
 \caption{ The phase $\Phi = (2 \pi -\phi(r)) \times N$ of the Wilson loop $W_C(r)$  
           is shown as a function of $r/N^{1/2}$ for the solutions with $N=4,7,10$.}
 \vbox{ \vskip -1.5 cm \hskip -0.3 cm \hbox{  \epsfxsize=450pt \hbox{\epsffile{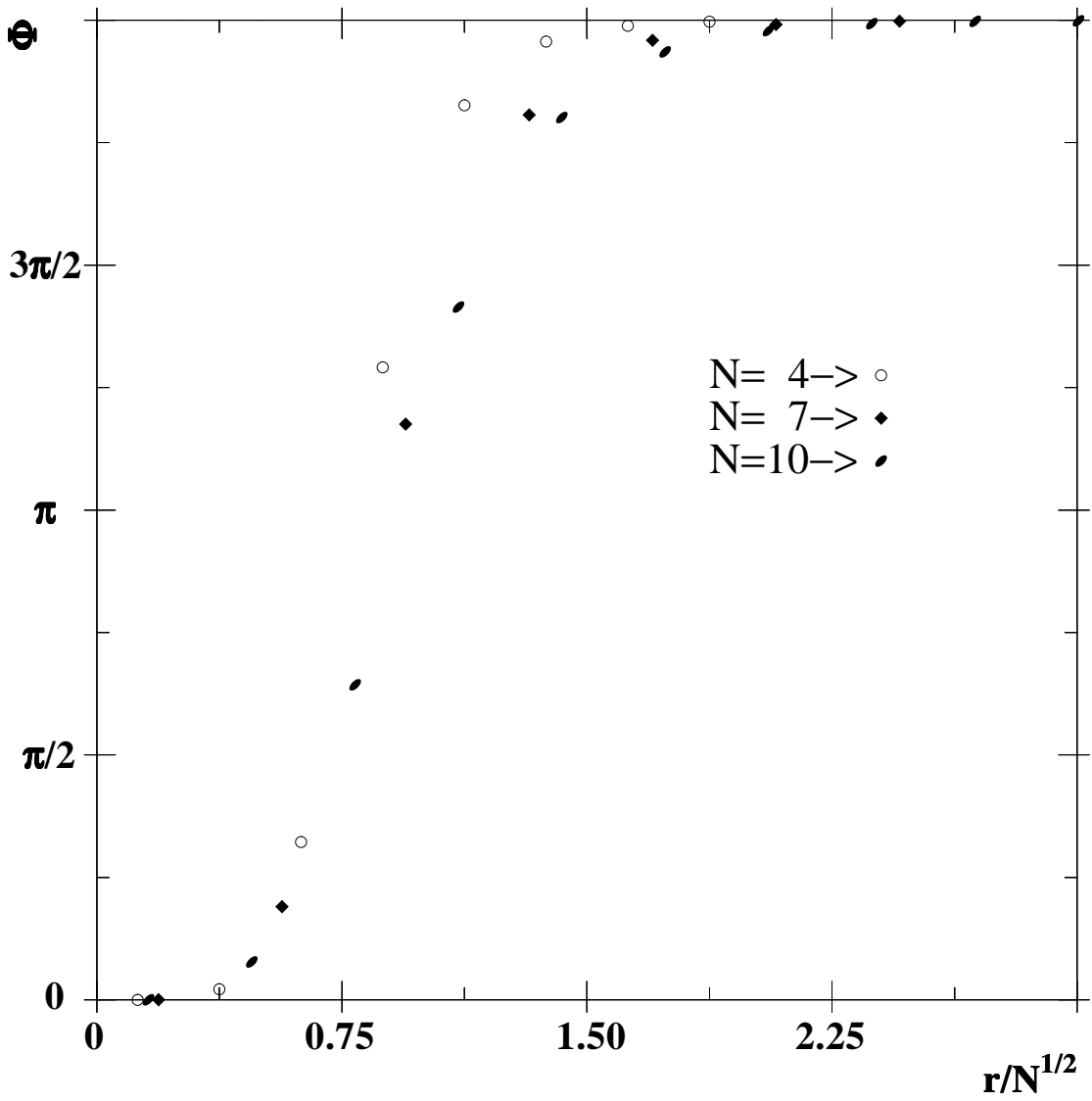} } } } 
\end{figure}

\end{document}